# The development status of the NIR Arm of the new SoXS instrument at the ESO/NTT telescope

F. Vitali*[a], M. Aliverti[b], G. Capasso[c], F. D'Alessio[a], M. Munari[d], M. Riva[b], S. Scuderi[e], R. Zanmar Sanchez[d], S. Campana[b], P. Schipani[c], R. Claudi[f], A. Baruffolo[f], S. Ben-Ami[i], F. Biondi[f], A. Brucalassi[n], R. Cosentino[h], D. Ricci[f], P. D'Avanzo[b], H. Kuncarayakti[j,k], A. Rubin[i], J. Achrén[l], J. A. Araiza-Duran[m,o], I. Arcavi[p], A. Bianco[b], R. Bruch[i], E. Cappellaro[f], M. Colapietro[c], M. Della Valle[c], M. De Pascale[f], R. Di Benedetto[d], S. D'Orsi[c], D. Fantinel[f], A. Gal-Yam[i], M. Genoni[b], M. Hernandez[h], O. Hershko[i], J. Kotilainen[j,k], M. Landoni[b], G. Li Causi[s], S. Mattila[k], G. Pignata[n,o], K. Radhakrishnan[f], M. Rappaport[i], B. Salasnich[f], S. Smartt[t], M. Stritzinger[u], E. Ventura[h], D. Young[t].

[a]INAF–Osservatorio Astronomico di Roma, Via Frascati 33, I-00078 M. Porzio Catone, Italy
[b]INAF–Osservatorio Astronomico di Brera, Via Bianchi 46, I-23807, Merate, Italy
[c]INAF–Osservatorio Astronomico di Capodimonte, Salita Moiariello 16, I-80131, Naples, Italy
[d]INAF–Osservatorio Astrofisico di Catania, Via S. Sofia 78, I-95123 Catania, Italy
[e]INAF–IASF Milano Via A. Corti, 12, I-20133 Milano, Italy
[f]INAF–Osservatorio Astronomico di Padova, Vicolo dell'Osservatorio 5, I-35122, Padua, Italy
[g]ESO, Karl Schwarzschild Strasse 2, D-85748, Garching bei München, Germany
[h]FGG-INAF, TNG, Rambla J.A. Fernandez Perez 7, E-38712 Brenã Baja (TF), Spain
[i]Weizmann Institute of Science, Herzl St 234, Rehovot, 7610001, Israel
[j]Finnish Centre for Astronomy with ESO (FINCA), FI-20014 University of Turku, Finland
[k]Tuorla Observatory, Dept. of Physics and Astronomy, FI-20014 University of Turku, Finland
[l]Incident Angle Oy, Capsiankatu 4 A 29, FI-20320 Turku, Finland
[m]Centro de Investigaciones en Optica A. C., 37150 León, Mexico
[n]Universidad Andres Bello, Avda. Republica 252, Santiago, Chile
[o]Millennium Institute of Astrophysics (MAS)
[p]Tel Aviv University, Department of Astrophysics, 69978 Tel Aviv, Israel
[q]Dark Cosmology Centre, Juliane Maries Vej 30, DK-2100 Copenhagen, Denmark
[r]Aboa Space Research Oy, Tierankatu 4B, FI-20520 Turku, Finland
[s]INAF - Istituto di Astrofisica e Planetologia Spaziali, Via Fosso del Cavaliere 100, 00133, Roma, Italy
[t]Astrophysics Research Centre, Queen's University Belfast, Belfast, BT7 1NN, UK
[u]Aarhus University, Ny Munkegade 120, D-8000 Aarhus

## ABSTRACT

We present here the development status of the NIR spectrograph of the Son Of X-Shooter (SOXS) instrument, for the ESO/NTT telescope at La Silla (Chile). SOXS is a R~4,500 mean resolution spectrograph, with a simultaneously coverage from about 0.35 to 2.00 µm. It will be mounted at the Nasmyth focus of the NTT. The two UV-VIS-NIR wavelength ranges will be covered by two separated arms. The NIR spectrograph is a fully cryogenic echelle-dispersed spectrograph, working in the range 0.80-2.00 µm, equipped with a Hawaii H2RG IR array from Teledyne. The whole spectrograph will be cooled down to about 150 K (but the array at 40 K), to lower the thermal background, and equipped with a thermal filter to block any thermal radiation above 2.0 µm. In this work, we will show the advanced phase of integration of the NIR spectrograph.

*fabrizio.vitali@inaf.it

# 1 INTRODUCTION

Son Of X-Shooter (SOXS) will be the new UV-VIS-NIR spectrograph for the ESO 3.6m New Technology Telescope in La Silla (Chile). SOXS in an international project, leaded by the Italian National Institute of Astrophysics (INAF). SOXS is a $R_S > 4,500$ resolution spectrograph, with a simultaneously coverage from about 0.35 to 2.00 μm. Its design foresees two separate high-efficiency spectrographs (UV–VIS and NIR) with a small overlapping range of about 80 nm for the spectral intercalibration. The NIR spectrograph is a fully cryogenic echelle-dispersed spectrograph, working in the range 0.80-2.00 μm. The optical design is based on a 4C echelle. The dispersion is obtained via a main disperser grating and three prisms as cross-dispersers. The NIR spectrum will be dispersed on 15 orders, with a minimum inter-order of ~10 px (in the blue part). The average throughput will be about 28%, including the telescope. The spectrograph will be equipped with a Hawaii H2RG IR array from Teledyne and controlled via the new NGC controller from ESO. It will be cooled down to about 150 K (but the array, cooled down to 40 K), to lower the thermal background, and equipped with a thermal filter to block any thermal radiation above 2.00 μm. The cryogenics will be operated via a Closed Cycle CryoCooler. The instrument is currently in the manufacturing and integration phase. The current emergency situation due to the COVID-19 has obviously delayed the SOXS schedule, however we expect to complete the procurement and shipping of all the parts of the NIR spectrograph within this year. The most affected work-package is the AIT, with serious restrictions both for our national laboratories and for the other international partners of the project. The overall description of the instrument can be found in [1] and [2]. This document describes the development status of the NIR arm of SOXS. Currently, the best guess is to have SOXS operative within 2022.

# 2 THE NIR SPECTROGRAPH

In this document we show the development status of the NIR spectrograph, whose full description can be found in [3]. In particular, we describe the status of the integration of the several opto-mechanical parts and devices, the cryogenics and the detector systems.

## 2.1 OPTICS

The optical design of the NIR arm of SOXS [4] is composed of a double pass collimator and a refractive camera; the dispersion is obtained via a main disperser grating and three prisms acting as cross-dispersers. The collimator receives a F/6.5 input from the Common Path and creates a collimated beam. The main disperser is a standard grating with 72 l/mm and a blaze angle of 44º, whereas three Cleartran prisms, used in double pass, provide the cross-dispersion. The camera, a completely transmissive system composed of three single lenses, re-images on the detector the focal plane produced by the second pass through the collimator and image the spectrum on 15 orders: Figure 1 shows the optical design of the NIR arm and a synthetic spectrum obtained through the SOXS instrument simulator.

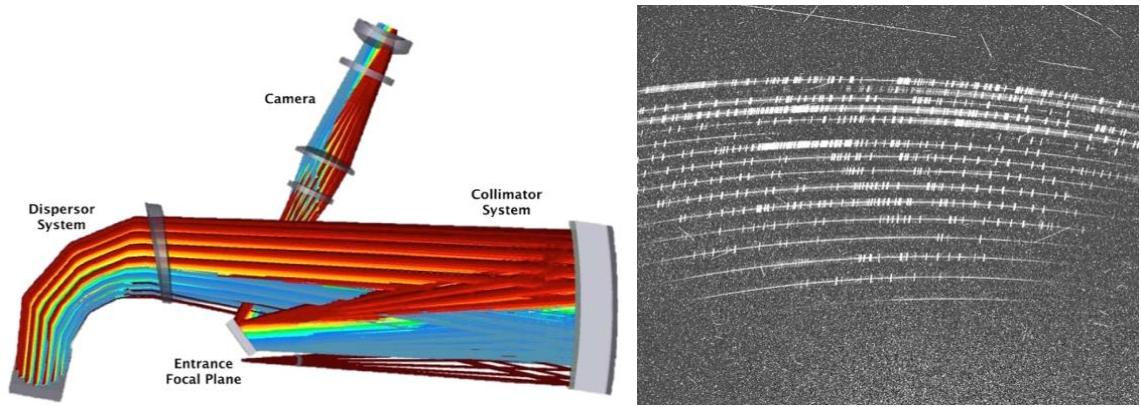

Figure 1: the optical design of the NIR arm (left) and the synthetic spectrum imaged on the NIR array (right).

NIR spectrograph optics procurement is almost completed, except for the aluminum mirrors and the corrector lens of the collimator. Some lenses systems have been already assembled in their mechanics.

Interferometric tests on NIR-Gratings for blaze angle check and mechanical characterization have been started. The blaze angle has been checked by minimizing fringes of the reflected beam from the NIR grating, illuminated with a collimated beam at 633nm (Zygo interferometer GPI-XP). In Figure 2 we show the grating mounted on its mounts and positioned on the alignment stage for the test. On the same set-up a cross check of the grating alignment has been done using Faro-Arm equipment.

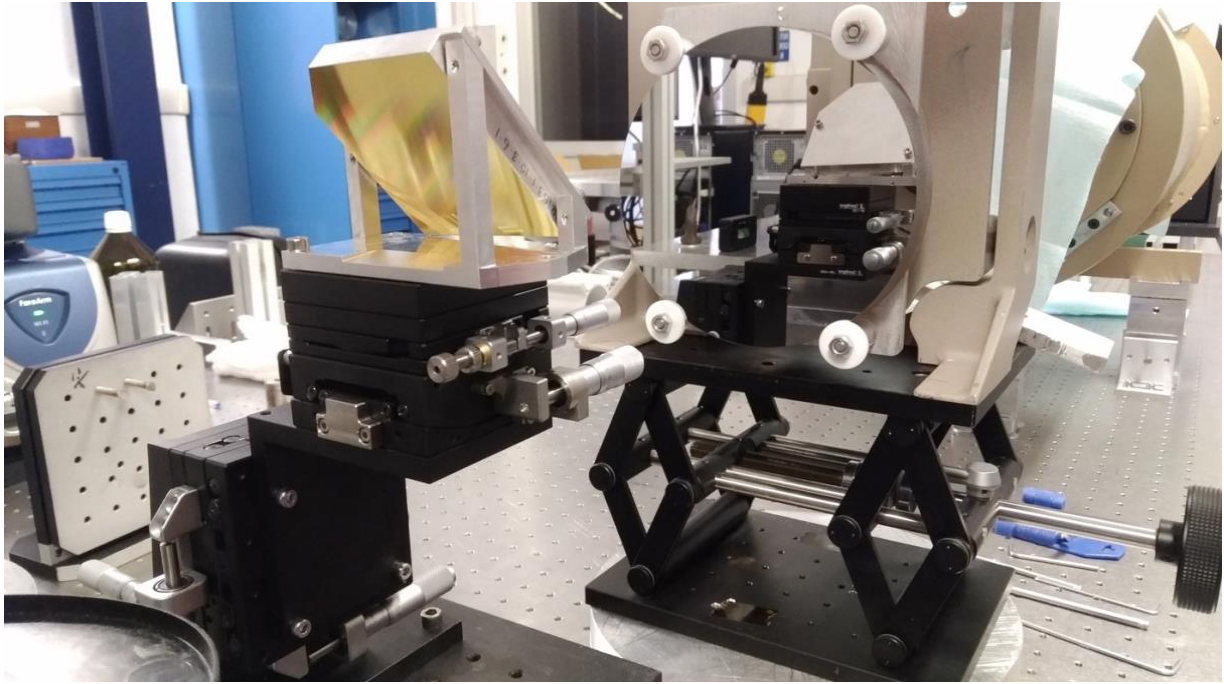

Figure 2: NIR grating mounted on its mount structure and positioned onto an alignment stage for interferometric test and mechanical characterization.

The Camera lenses and cross-disperser prisms have been tested to check throughput performance with a measurement set-up made by a scanning monochromator (Newport Cornerstone-130), iris (THORLABS ID25/M) and photodiode sensors (THORLABS S132C Ge-Photodiode), see Figure 3.

The Al mirrors are currently under final coating. They are expected to be delivered within January 2021. In Figure 4 the collimator mirror is showed after the first milling at Astron-Nova (left) and the small mirrors , after the black coating, to reduce stray light (right). All process steps have been followed and the relevant mechanical tolerances have been confirmed by CMM measurements.

We tested all the delivered optics in our laboratories and derived the efficiency of the main four optical systems (cross-disperser, grating and camera), whose efficiency curves are reported in Figure 5 (left), resulting in a total efficiency curve for the NIR spectrograph optics (no detector QE) as showed in Figure 5 (right). For the collimator, we used the theoretical transmission curve.

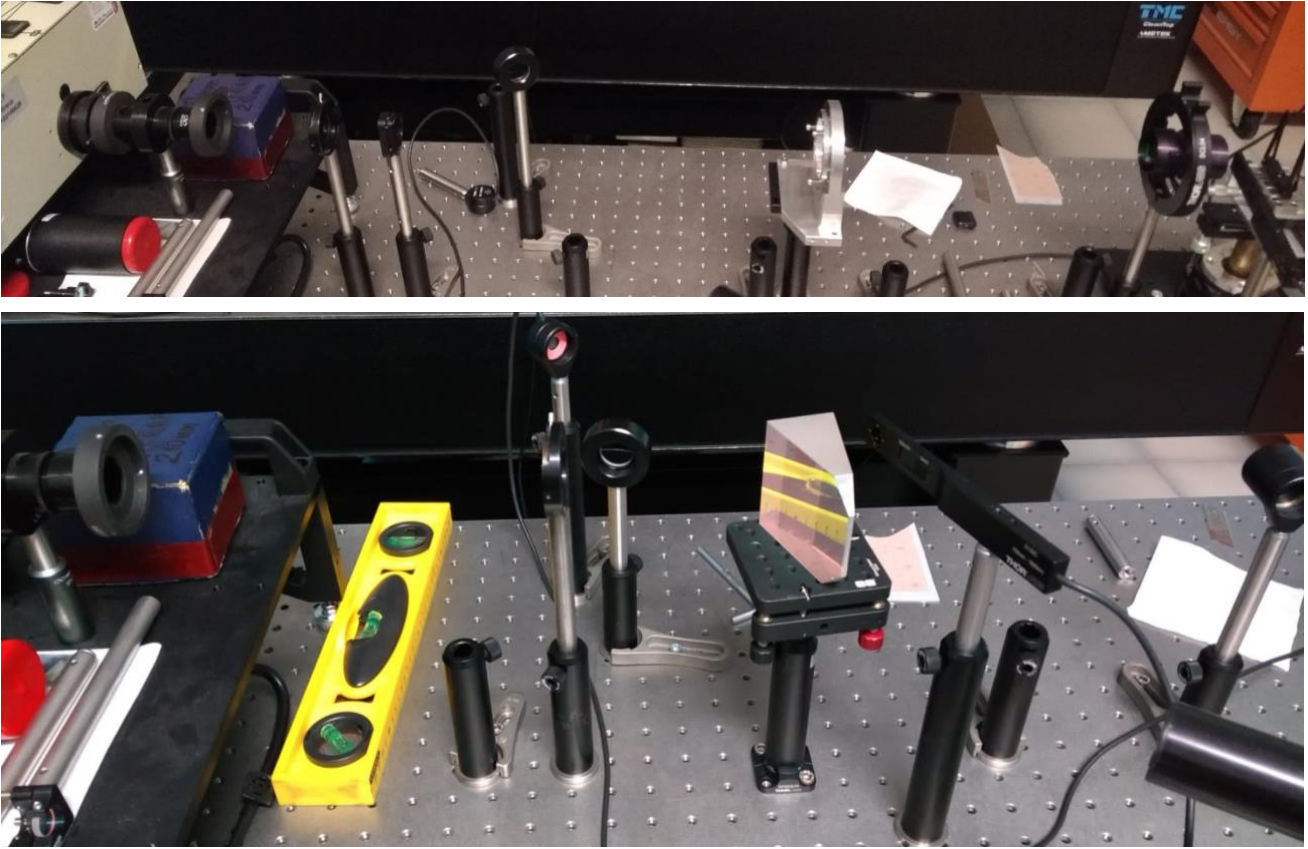

Figure 3: Laboratory measurement set-up for NIR camera lenses (above) and cross-disperser prisms (below) transmission tests.

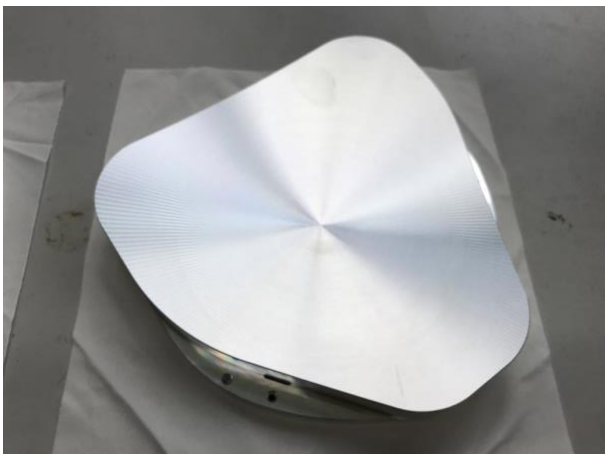 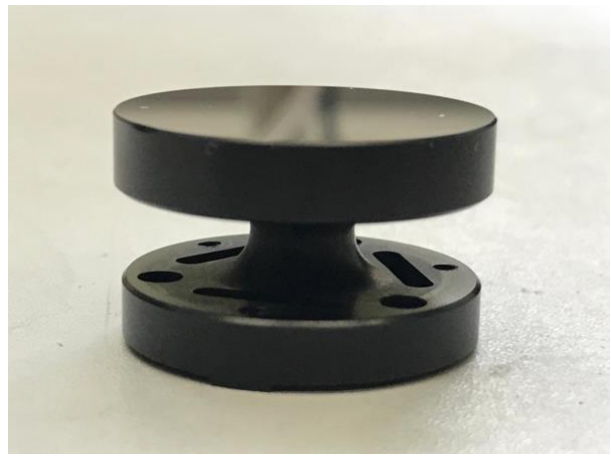

Figure 4: Collimator mirror at Astron-Nova after first milling (left) and the small mirror, after the black coating, to reduce stray light, before the final gold coating (right).

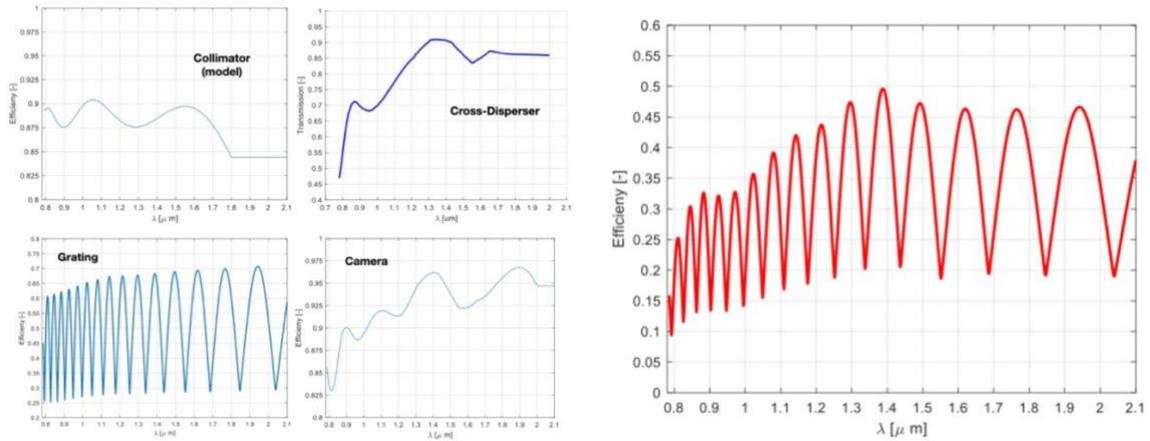

Figure 5: Efficiency curves for the collimator (model), cross-disperser, grating and camera lens systems (left) and the total efficiency curve (with no detector QE).

## 2.2 MECHANICS

An overall view of the NIR vacuum vessel is showed in Figure 6, where all the cryo-vacuum devices, the opto-mechanical subsystems and the detector system are labelled. A full description of the NIR spectrograph mechanics can be found in [5]

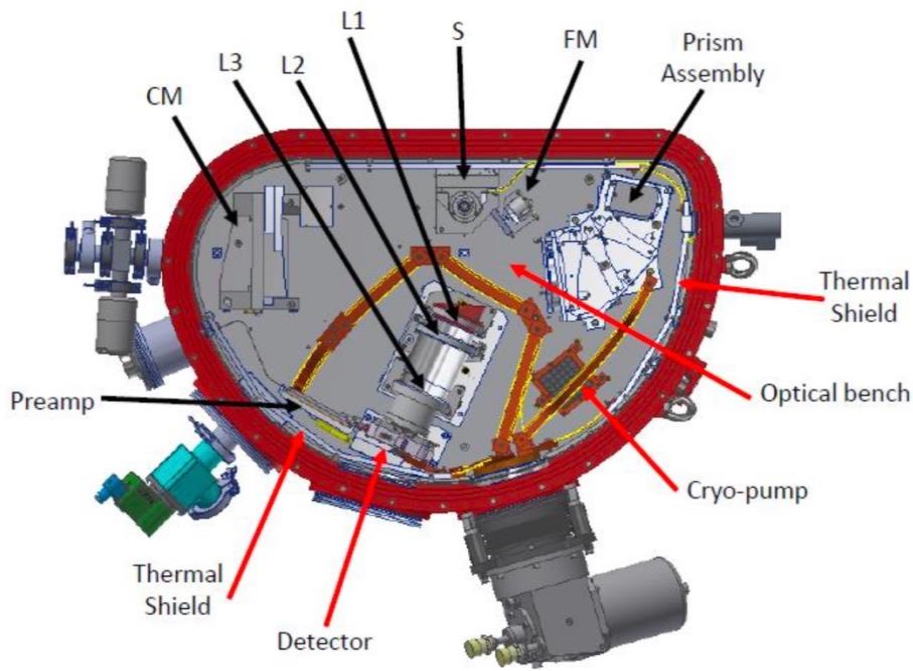

Figure 6: the internal design of the NIR vessel.

By the end of July 2020, the vacuum vessel was delivered and preliminary vacuum tests were performed. Figure 7-left shows the D-shaped vacuum vessel, that will be interfaced to the telescope interface flange, through a set of kinematic mounts. Most of the NIR spectrograph opto-mechanics elements have been assembled, to perform the first tests on the whole system. In particular, in Figure 7-right you can see the camera (L1, L2 and L3 in Figure 6), the cross-disperser (with the three prism holders) and the entrance optics systems (S and FM in Figure 6).

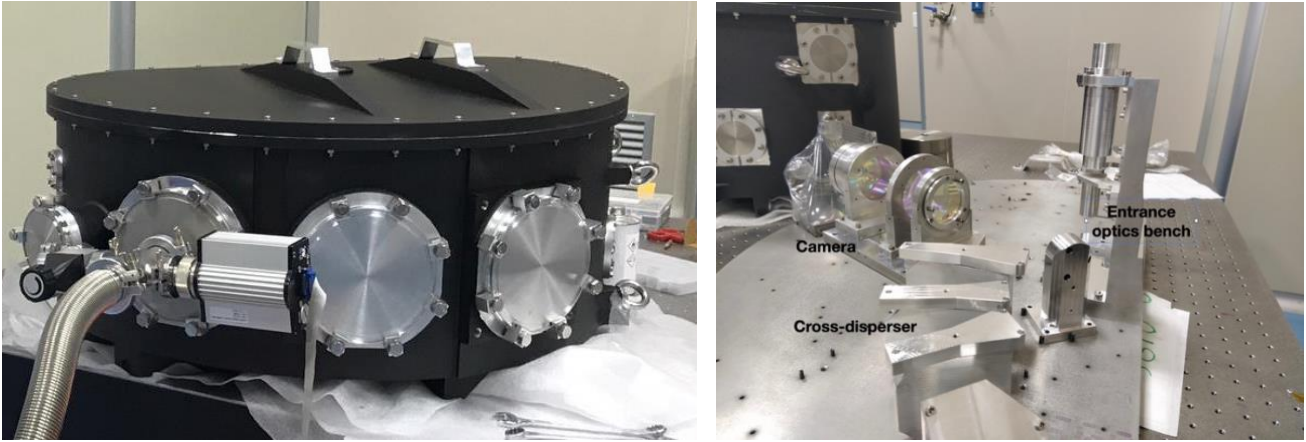

Figure 7: the NIR vacuum cryostat (left) and the cold plate, where some systems are assembled (right).

In July 2020, the supports of the NIR camera were received. The parts have been accepted, cleaned and placed into a vacuum vessel with some dummy lenses installed (cheap flat windows with the same material and interface as the actual ones). Resistance tests were performed at high temperature (90°C), while the tests at low temperature (150K) are going to be executed soon.

## 2.3 CRYOGENICS

Functional tests for the NIR cryo-vacuum system have been performed using a dummy vessel adapted to simulate the NIR spectrograph and in manual mode, i.e., with no control through Siemens PLC (Figure 8, left). The laboratory set up is being used to calibrate the different temperature sensors (diodes DT670 and platinum resistances PT100) in their working temperature range.

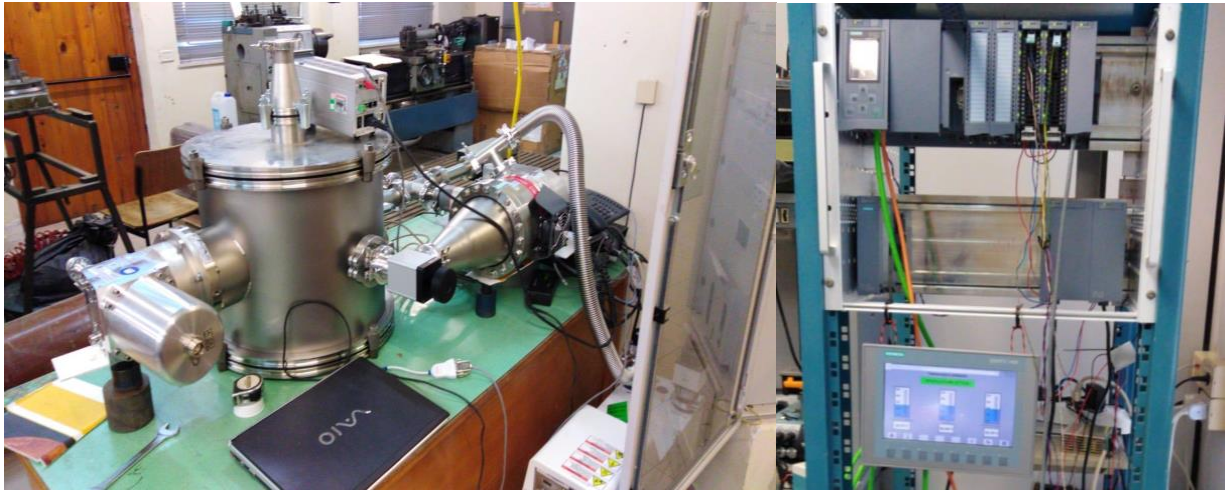

Figure 8: Vacuum & Cryogenics activity in INAF-OA Catania (left) and the Control Electronics rack (right).

Finally, the Siemens modules were all integrated in a rack (Figure 8, right). Communication tests with the various sensors (pressure gauges and temperature probes) and control tests on vacuum pumps and valves have been performed successfully.

Furthermore, the capabilities of the cooling head (Leybold Coolpower 250 MD) have been tested against various thermal loads (Figure 9).

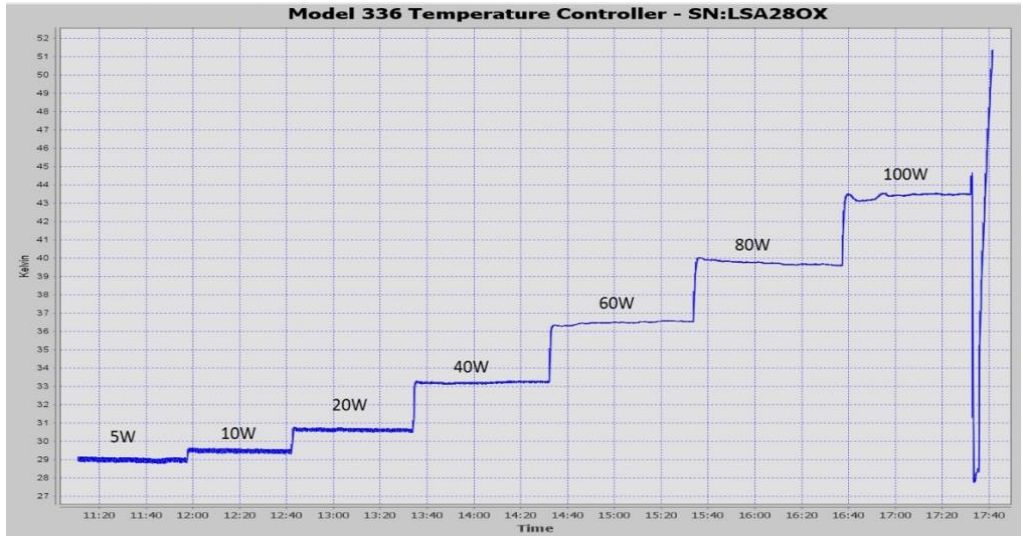

Figure 9. Cooling head response against various thermal loads.

## 2.4 NIR DETECTOR SYSTEM

The detector system includes the H2RG NIR array and the NGC control electronics.

The Hawaii H2RG NIR array was delivered by Teledyne on September 2020 at the INAF-OAB premises. Teledyne provided us with a full set of tests on the scientific array, all the performances were well in accordance with the required specifications, as reported in Figure 10, in particular in terms of dark current (0 e$^-$/s), noise (about 11 e- rms in CDS and 3.62 e- rms in Fowler-16 sample), persistence and operability.

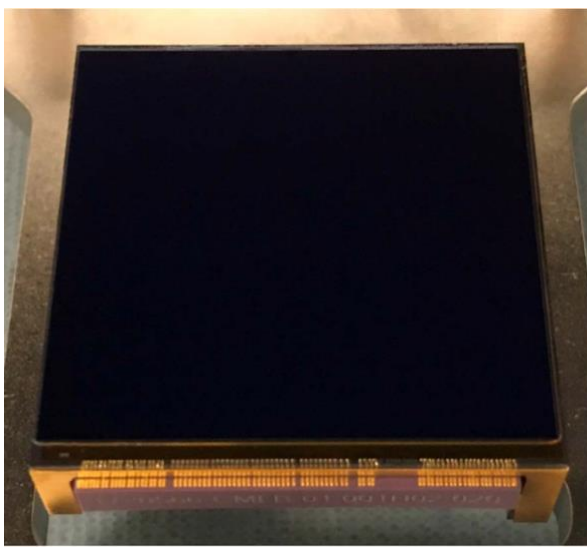

Figure 10: the SOXS H2RG array from Teledyne (left) and the performance summary (right).

To evaluate the persistence effect, Teledyne reported the dark current as a function of time for 6 hours after the illumination exposure (Figure 11). Each dark current point is calculated from a 300 second integration ramp. The initial point is from an integration ramp prior to the exposure to light and the final point is the one taken the next morning after illumination. In Figure 11, left, the extremely good flatness and cosmetics of the array can be appreciated.

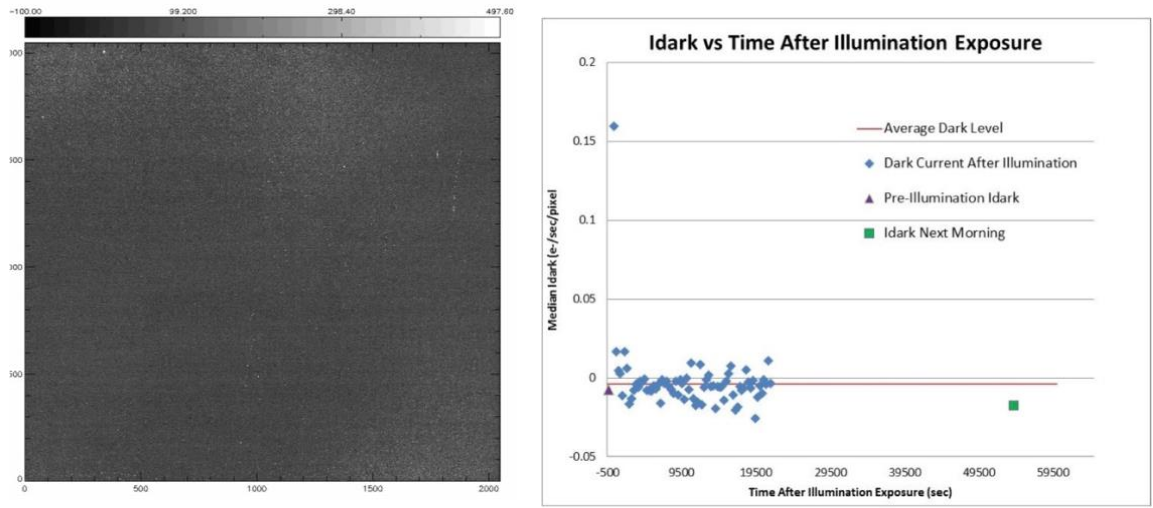

Figure 11: the dark current frame (left) and as a function of time, after an illumination exposure (right).

Operability is very high, bad pixels are calculated based upon the following values: QE ≥ 35%, Idark ≤0.5 e-/s and CDS Noise ≤35.0 e-. A cluster is defined as 50 or more contiguous pixels.

Meanwhile, the components for the NGC controller system were manufactured, shipped to ESO Garching and assembled in the ESO detector lab. The electronic system is quite ready for the shipping to the INAF-Roma (OAR) in Monte Porzio Catone. The delivery is expected by the end of December 2020.

At the end of 2019, a training period was carried out at ESO's detector lab, where we verified the characteristics of the new ESO NGC controller.

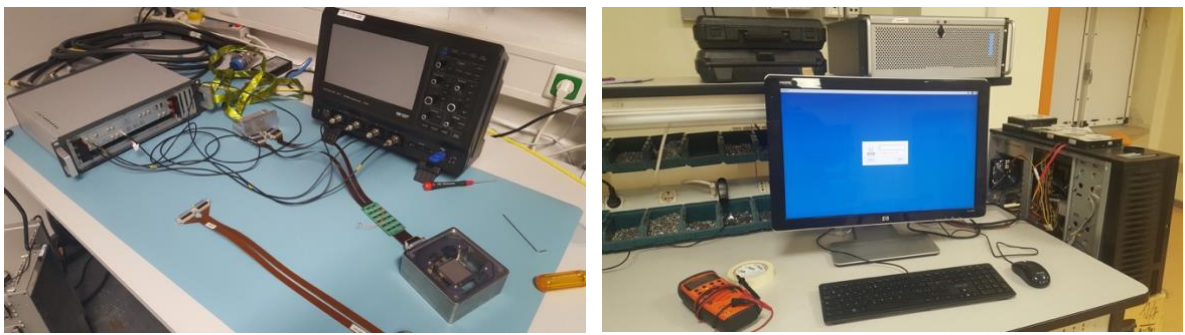

Figure 12: NGC system at ESO-Garching labs (left) and the LLCU at INAF-OAR lab (right).

We worked for two weeks with the LLCU control PC for SOXS, ready to manage the HAWAII 2RG controller. Then the LLCU was immediately sent to the laboratories of the INAF-OAR (Figure 12). The new NGC controller for SOXS infrared arm is ready and is still at ESO's detector labs in Garching. At the beginning of 2021, we expect to have the whole electronic system in the INAF-OAR lab in Monte Porzio Catone, for the final electronic tests and we foresee to send it to INAF-OAB in Merate, for the cryostat assembly and cold tests of NIR cryostat.

## 2.5 CONTROL SYSTEM

A detailed description of the control system can be found in [6]. A dedicated Micronix controller controls the linear cryogenic stage inside the NIR vacuum vessel. It has been integrated inside the Instrument Control Electronics at the Capodimonte Observatory electronic lab (INAF-OACN) in Naples, Italy (Figure 13). See [7] for further details.

Testing of the controller and of the NIR slit exchanger have been carried in Padua Observatory (INAF-OAPD), see [8] for details.

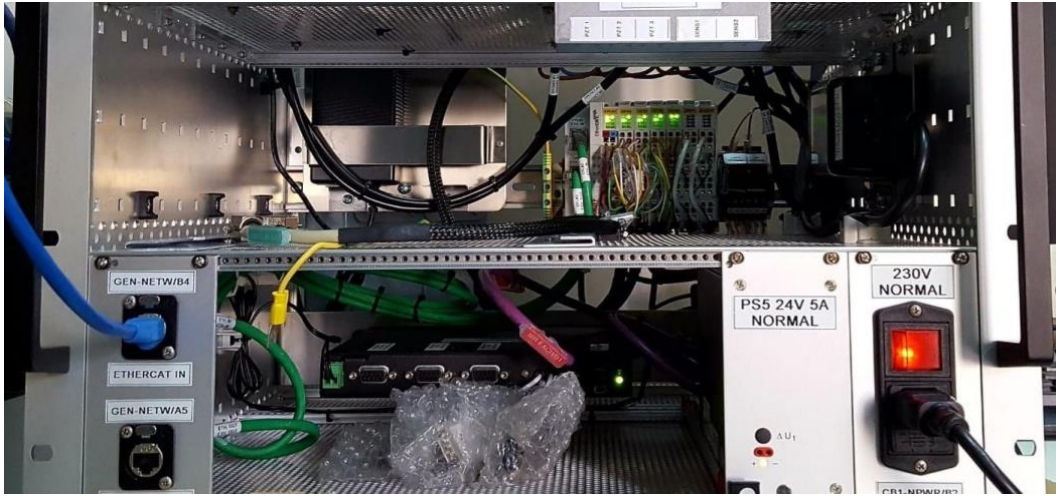

Figure 13: EtherCAT to RS-485 connection and NIR slit controller at INAF-OACN lab

## 3 CONCLUSION

We described the development status of the NIR arm of the SOXS spectrograph, to be installed in the Nasmyth room of the ESO-NTT telescope, in La Silla (Chile), in 2022. Almost all the opto-mechanical parts have been delivered and are currently under AIT at the INAF laboratories. Once each subsystem is fully mounted, aligned and tested, these will be shipped to the INAF-OAP in Padova, for the final AIT in Europe.

## REFERENCES

[1] Schipani, P. et al., "SOXS: a wide band spectrograph to follow up transients", Proc. SPIE 10702, 107020F (2018).
[2] Schipani P. et al., "Development status of the SOXS spectrograph for the ESO-NTT telescope", *Proc. SPIE* 11447 (2020).
[3] Vitali, F. et al., "The NIR spectrograph for the new SOXS instrument at the NTT," Proc. SPIE 10702, 1070228 (2018).
[4] R. Zanmar Sanchez, et al., "Optical design of the SOXS spectrograph for ESO NTT", Proc. SPIE 10702, (2018).
[5] M. Aliverti, et al., "The mechanical design of SOXS for the NTT", Proc. SPIE 10702, (2018).
[6] G. Capasso, et al., "SOXS control electronics design", Proc. SPIE 10707, (2018).
[7] Colapietro, M. et al., "Progress and tests on the Instrument Control Electronics for SOXS", *Proc. SPIE* 11452-89 (2020).
[8] Ricci, D. et al., "Development status of the SOXS instrument control software", *Proc. SPIE* 11452 (2020).